\documentclass[twocolumn,english,showpacs,preprintnumbers,amsmath,amssymb,prb,superscriptaddress]{revtex4-1}

\usepackage[T1]{fontenc}
\usepackage[latin2]{inputenc}
\usepackage{color}
\usepackage{array}
\usepackage{longtable}
\usepackage{float}
\usepackage{booktabs}
\usepackage{textcomp}
\usepackage{amsmath}
\usepackage{graphicx}
\usepackage{amssymb}
\usepackage{multirow}

\begin{document}

\title{Magnetic properties of the low-dimensional spin-1/2 magnet $\alpha$-Cu$_{2}$As$_{2}$O$_{7}$}

\author{Y. C. Arango}

\affiliation{Leibniz Institute for Solid State and Materials Research IFW Dresden,
D-01171 Dresden, PO BOX 270116, Germany}

\author{E. Vavilova}

\affiliation{Leibniz Institute for Solid State and Materials Research IFW Dresden,
D-01171 Dresden, PO BOX 270116, Germany}

\affiliation{Zavoisky Physical Technical Institute of the Russian Academy of Sciences,
420029, Kazan, Russia}

\author{M. Abdel-Hafiez }

\affiliation{Leibniz Institute for Solid State and Materials Research IFW Dresden,
D-01171 Dresden, PO BOX 270116, Germany}

\author{O. Janson}

\author{A. A. Tsirlin}

\author{H. Rosner}

\affiliation{Max Planck Institute for Chemical Physics of Solids, D-01187 Dresden,
Germany}

\author{S.-L. Drechsler}

\affiliation{Leibniz Institute for Solid State and Materials Research IFW Dresden,
D-01171 Dresden, PO BOX 270116, Germany}

\author{M. Weil}

\affiliation{Institute for Chemical Technologies and Analytics, Vienna University
of Technology, A-1060 Vienna, Austria}

\author{G. N{{{{\'e}}}}nert}

\affiliation{Institut Laue-Langevin, Bo{{{{\^\i}}}}te Postale 156, 38042 Grenoble
Cedex 9, France}

\author{R. Klingeler}

\affiliation{Kirchhoff Institute for Physics, University of Heidelberg, D-69120
Heidelberg, Germany}

\author{O. Volkova}

\author{A. Vasiliev}

\affiliation{Low Temperature Physics Department, Moscow State University, Moscow
119991, Russia}

\author{V. Kataev}


\author{B. B{{\"u}}chner}

\affiliation{Leibniz Institute for Solid State and Materials Research IFW Dresden,
D-01171 Dresden, PO BOX 270116, Germany}

\date{\today}
\begin{abstract}
 In this work we study the interplay between the crystal
structure and magnetism of the pyroarsenate $\alpha$-Cu$_{2}$As$_{2}$O$_{7}$ by means of magnetization, heat capacity, electron spin resonance and
nuclear magnetic resonance measurements as well as density functional theory (DFT) calculations and quantum Monte Carlo (QMC) simulations. The data
reveal that the magnetic Cu-O chains in the crystal structure represent a realization of a quasi-one dimensional (1D) coupled alternating spin-1/2
Heisenberg chain model with relevant pathways through non-magnetic AsO$_{4}$ tetrahedra. Owing to residual 3D interactions antiferromagnetic long
range ordering at $T_{\textrm{N}}\simeq10$\,K takes place. Application of external magnetic field \textit{B} along the magnetically easy axis induces
the transition to a spin-flop phase at \textit{B$_{\textrm{SF}}$} $\sim$1.7\,T (2 K). The experimental data suggest that substantial quantum spin
fluctuations take place at low magnetic fields in the ordered state. DFT calculations confirm the quasi-one-dimensional nature of the spin lattice,
with the leading coupling $J_{1}$ within the structural dimers. QMC fits to the magnetic susceptibility evaluate $J_{1}=164$\,K, the weaker
intrachain coupling $J_{1}^{\prime}/J_{1} = 0.55$, and the effective interchain coupling $J_{\textrm{ic1}}/J_{1} = 0.20$.
\end{abstract}


\maketitle

\section{introduction}

Crystals of composition \textsl{X}$_{2}$\textsl{Y}$_{2}$\textsl{O}$_{7}$ tend to adopt different structures in which the \textit{Y} atom may be
octahedrally or tetrahedrally coordinated \cite{Wyckoff}. These systems have been found to exhibit interesting magnetic properties ranging from a
geometry induced spin frustration phenomena when \textit{X} and \textit{Y} are magnetic metal ions (see, e.g., Ref. \cite{Gardner}), to a peculiar
interplay between structure and magnetism in those compounds where \textit{Y} is not magnetic (see, e.g., Refs. \cite{Pommer2003,Tsirlin2010}). Co,
Ni, and Cu diphosphates and divanadates are interesting examples of those latter compounds. In their crystal structure two\textit{ Y}O$_{4}$
tetrahedra sharing an oxygen atom usually undergo a transition from a staggered ($180^{\circ}$) to a bent \textit{Y}\,-\,O\,-\,\textit{Y}
configuration (from thortveitite to dichromate type structure) with the decrease of temperature. A recent comparative band-structure study of copper
(II) based pyrocompounds Cu$_{2}$\textit{Y}$_{2}$O$_{7}$ (\textit{Y}=P, As, V) \cite{Janson} shows that despite having the same basic structural
motif these materials represent completely different magnetic models. Here the change of the non-magnetic central atom \textit{Y} in the tetrahedral
anion groups \textit{Y}O$_{4}$, e.g. from \textit{Y} = P to V, turns out to be crucial in determining the effective interchain exchange pathways
depending on the electronic state (\textit{p}- or \textit{d}-type) of the central \textit{Y} cation. Moreover, significant variations in the
intrachain exchange coupling evidence the subtle effect of the Cu-O-Cu bond angle albeit the basic structural dimerization of Cu$_{2}$O$_{6}$
plaquettes is present in all these isostructural compounds. Thus, according to theoretical predictions and experimental data available so far, one
can expect that further change of the \textit{Y} cation, might yield a completely new magnetic behavior in this series of materials. The validation
of this conjecture would be of a specific importance for the development and verification of new magnetic models of low-dimensional Heisenberg
quantum magnets which since decades stand in the focus of intense experimental and theoretical research in condensed matter physics \cite{quantmag}.

\begin{figure}
\begin{centering}
\includegraphics[angle=-90,width=\columnwidth]{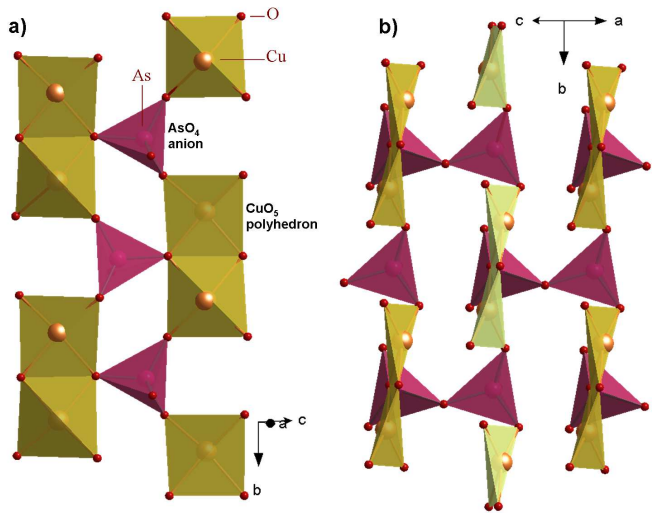}
\par\end{centering}

\caption{(Color online) The crystal structure of $\alpha$-Cu$_{2}$As$_{2}$O$_{7}$: (a) - edge-sharing Cu$_{2}$O$_{6}$ plaquettes form structural
dimer units along the \textit{b}-axis. The dimers are linked by AsO$_{4}$ anion groups forming a plane of coupled dimer chains; (b) - side view of
the chain planes linked with interjacent As$_{2}$O$_{7}$ anions.}

\label{fig:crys_struc}
\end{figure}

Unlike pyrovanadates and even pyrophosphates, the experimental magnetic study of transition metal pyroarsenates, \textsl{X}$_{2}$As$_{2}$O$_{7}$
(\textsl{X} = Cu, Ni, Co, Mn) is scarce. Only for \textsl{X}$_{2}$As$_{2}$O$_{7}$ with \textsl{X} = Ni, Co, Mn, early bulk magnetic susceptibility
measurements and low-temperature neutron powder diffraction were reported by Buckley \textsl{et al.}, \cite{Buckley1995} whereas there seem to be no
experimental results on the copper (II)-based Cu$_{2}$As$_{2}$O$_{7}$ pyroarsenate. Therefore, in view of the expected sensitivity of the magnetic
behavior to the type of the \textit{Y} cation we have carried out a detailed study of magnetic properties of single crystals of
$\alpha$-Cu$_{2}$As$_{2}$O$_{7}$ with different experimental techniques such as magnetization, heat capacity, electron spin resonance (ESR) and
nuclear magnetic resonance (NMR) measurements, combined with density functional theory (DFT) calculations and quantum Monte Carlo (QMC) simulations.
Indeed, we find that in contrast to other related isostructural compounds $\alpha$-Cu$_{2}$As$_{2}$O$_{7}$ represents an experimental realization of
coupled spin-1/2 alternating antiferromagentic (AFM) Heisenberg chains where the AsO$_{4}$ anion groups play an important role for the interchain
magnetic exchange. The quasi-one dimensional magnetic behavior with short-range AFM correlations is dominating over a wide temperature range. At low
temperatures, however, secondary magnetic exchange interactions become relevant, too, and three-dimensional long-range magnetic order evolves below
$T_{{\rm \textrm{N}}}\simeq10$ K. In the magnetically ordered state, the application of moderate magnetic fields $B\simeq1.7$ T parallel to the
crystallographic $b$-axis yields a field induced reorientation of the spin lattice. Further increase of the magnetic field progressively suppresses
spin fluctuations which are evident in the thermodynamic and magnetic resonance response at small fields.

\section{Crystal structure}

Synthesis and characterization of $\alpha$-Cu$_{2}$As$_{2}$O$_{7}$ single crystals have been reported in Ref. \cite{Weil}. The
$\alpha$-Cu$_{2}$As$_{2}$O$_{7}$ modification crystallizes isotypically with $\alpha$-Cu$_{2}$P$_{2}$O$_{7}$ and $\beta$-Cu$_{2}$V$_{2}$O$_{7}$
structures, belonging to the monoclinic space group \textsl{C}2/\textsl{c} with \textsl{a} = 7.237(3), \textsl{b} = 8.2557(17), \textsl{c} = 9.780(3)
\r{A}, $\beta$= 111.03(2)\textdegree{}.
%
%
%
The major structural feature of $\alpha$-Cu$_{2}$As$_{2}$O$_{7}$ is the occurrence of non-planar Cu$_{2}$O$_{6}$ plaquettes (Cu-O-Cu bond of
101.68(8)\textdegree{} and Cu-O bond lengths of 1.92-2.00 $\textrm{\AA}$) forming structural dimers along the crystallographic \textsl{b} direction
\cite{Weil} (Fig. \ref{fig:crys_struc}a). The AsO$_{4}$ tetrahedra link these Cu-O-Cu dimers in a chain along the \textsl{b}-axis and also
interconnect the neighboring chains in the chain plane (Fig. \ref{fig:crys_struc}a). The chain planes are linked by interjacent pyroarsenate anions
As$_{2}$O$_{7}$ in a bent As-O-As configuration of 145.9(2)\textdegree{}. The 4-fold oxygen in-plane coordination of Cu is completed by the fifth
oxygen atom from the neighboring plane with a much larger bond length of 2.27 $\textrm{\AA}$ which therefore should only weakly contribute to magnetism.

\section{results and discussion}

\subsection{Magnetization}

\begin{figure}
\begin{centering}
\includegraphics[scale=0.31]{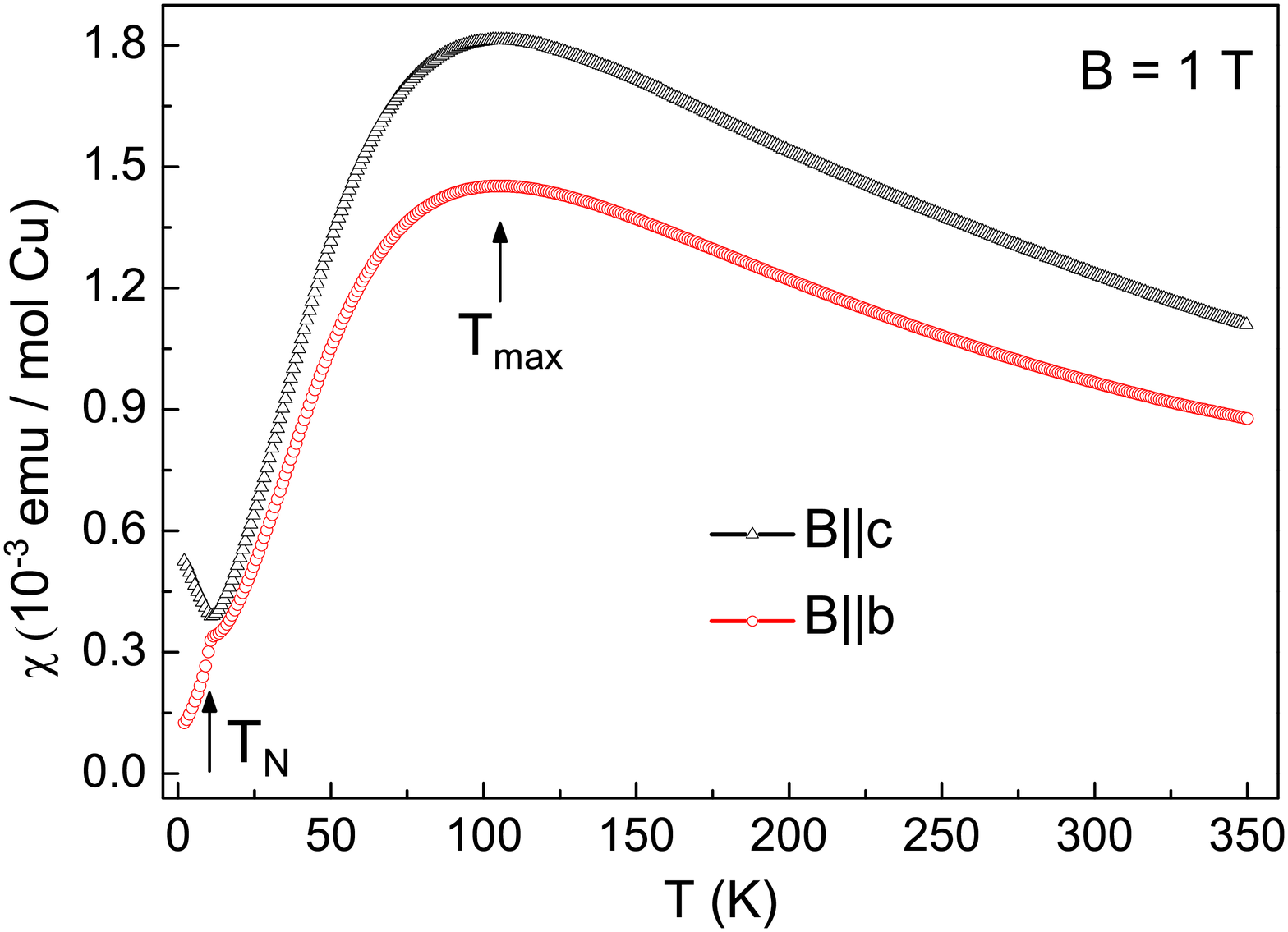}
\par\end{centering}

\caption{(Color online) Magnetic susceptibility $\chi=M/B$ of a $\alpha$-Cu$_{2}$As$_{2}$O$_{7}$ single crystal measured at $B=1$\,T applied along
the crystallographic $b$- and $c$-axis, respectively.}

\label{fig:Susc}
\end{figure}

 Magnetization measurements on a single crystal of $\alpha$-Cu$_{2}$As$_{2}$O$_{7}$
were carried out in a VSM SQUID magnetometer (Quantum Design) applying the magnetic fields parallel (\textsl{$B\Vert b$}-axis) and perpendicular
(\textsl{$B\Vert c$}-axis) to the propagation direction of the Cu$_{2}$O$_{6}$ plaquettes, respectively. Fig. \ref{fig:Susc} shows the static
magnetic susceptibility $\chi = M/B$ in the temperature range 2\,-\,350\,K at \textsl{$B=1$} T. For both field directions, the susceptibility data
exhibit a characteristic broad maximum at about $T\sim 105$\,K which is typically associated with 1D short-range magnetic correlations due to
predominant intra-chain magnetic interactions. This value suggests thus a rough estimate of the expected energy scale of the leading exchange
constants. At a much smaller temperature $T_{{\rm N}}\simeq10$ K, there is an upturn in $\chi$(\textsl{T})$\parallel$\textsl{c} and a concomitant
kink in $\chi$(\textsl{T})$\parallel$\textsl{b} evidencing the onset of long-range AFM spin order (see also figures \ref{fig:Heat_cap} and
\ref{NMRT1}). Such a feature and the rapid decrease of $\chi(T)\Vert b$ down to $T=2$ K imply the crystallographic \textit{b}-axis being the
magnetically \textit{easy}
axis. %
\begin{figure}
\begin{centering}
\includegraphics[scale=0.31]{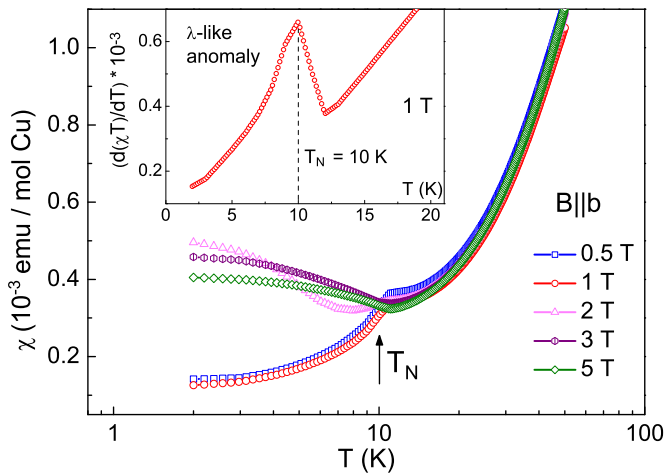}
\par\end{centering}

\caption{(Color online) Temperature dependence of the magnetic susceptibility $\chi(T)$. Inset: The pronounced anomaly in the Fisher's heat capacity
$\partial$($\chi T$)/$\partial T$ at $B=1$\,T along the $b$-axis implies long-range magnetic order below $T_{{\rm N}}\simeq10$ K. }

\label{fig:Susc_anom}
\end{figure}

\begin{figure}
\begin{centering}
\includegraphics[width=0.95\columnwidth]{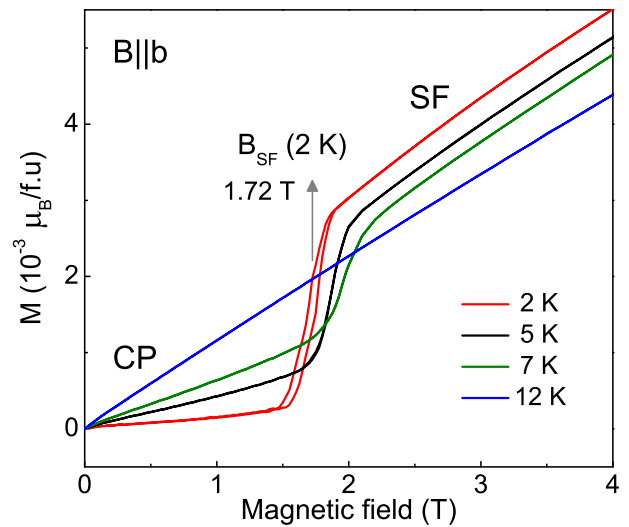}
\par\end{centering}

\caption{(Color online) Magnetic field $B$ dependence of the magnetization $M$ for $B\parallel b$-axis at different temperatures slightly above and
below $T_{\textrm{N}}$. CP and SF denote the presumably collinear phase in the low-field regime and the spin-flop phase at high magnetic fields. The
spin-flop transition at $B_{\textrm{SF}}\simeq1.7$\,T ($T=2$\,K) is associated with the magnetic anisotropy in the system.}

\label{spinflop}
\end{figure}

Further $\chi(T)\Vert b$ data at various constant applied magnetic fields are shown in Fig. \ref{fig:Susc_anom}. At low field $B<2$ T, the magnetic
specific heat which is proportional to $\partial$($\chi T$)/$\partial T$ exhibits a pronounced $\lambda$-like anomaly (Fisher's heat
capacity)\cite{Fisher1960,Fisher1962} corroborating the scenario of long range magnetic ordering at $T_{{\rm N}}=10$ K (Fig. \ref{fig:Susc_anom},
inset). Interestingly, the kink in $\chi(T)$ at $T_{{\rm N}}$ significantly changes its shape upon application of magnetic field $B>2$ T. The
decrease of $\chi(T)$ below $T_{{\rm N}}$ turns into a steep upturn at higher fields (Fig. \ref{fig:Susc_anom}). This transformation should be
indicative of an additional magnetic phase transition at a critical field in between these values and the development of a field-induced magnetic
phase. Such a behavior is indeed clearly visible in the magnetic field dependence of the magnetization $M$ vs. $B\|b$ (Fig. \ref{spinflop}).

Field dependent measurements of magnetization for \textsl{$B\Vert b$}-axis in Fig. \ref{spinflop} give evidence of a first-order spin-flop transition
(SF) below \textsl{$T_{\textrm{N}}$} at \textit{$B_{\textrm{SF}}\simeq1.7$ }T (2 K), in agreement with the $\chi(T)$ data. The spin-flop arises due
to the gain in magnetic anisotropy energy of the system above \textit{$B_{\textrm{SF}}$} by considering the AF collinear arrangement of Cu$^{2+}$
spins parallel to the \textsl{b}-axis at lower fields. The magnetization \textit{$M(B)$} data for \textit{$B\Vert c$}-axis in the field range up to 7
T (not shown) are consistent with a magnetically \textit{hard}
\textit{c}-axis, where no anomaly is observed below 12 K. %

\subsection{Heat capacity}

\begin{figure}
\begin{centering}
\includegraphics[width=0.85\columnwidth]{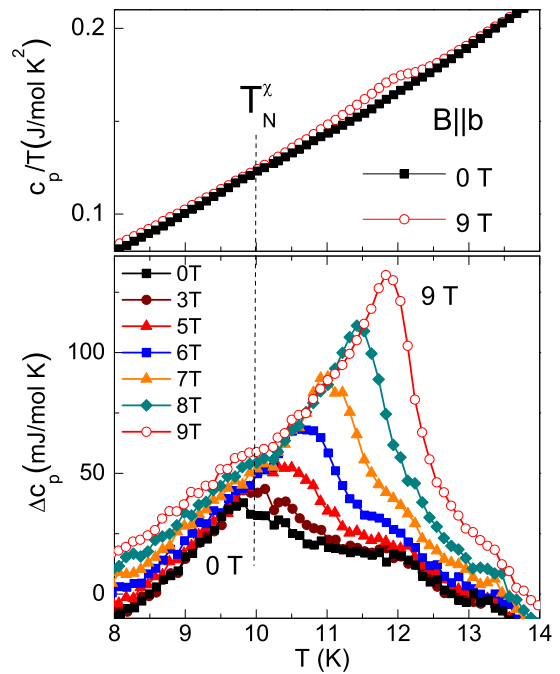}
\par\end{centering}

\caption{(Color online) Upper panel: Temperature dependence of the specific heat capacity $c_{{\rm p}}$ at $B=0$ and 9\,T; lower panel: Experimental
data in different magnetic fields $B\|b$-axis after subtraction of an arbitrary polynomial background function $f$($T$) = a$T$+b$T^{3}$. $T_{{\rm
N}}^{\chi}$ indicates the temperature of the anomaly in $\partial$($\chi T$)/$\partial T$ at small fields (cf. Fig.~\ref{fig:Susc_anom}).}

\label{fig:Heat_cap}
\end{figure}

Fig. \ref{fig:Heat_cap} shows the molar specific heat at $B=0$ and in different external magnetic fields applied parallel to the \textit{b}-axis. At
$B=0$, the as measured data do not exhibit a pronounced anomaly at $T_{\textrm{N}}$ despite the marked peak in the $\partial$($\chi T$)/$\partial T$
(Fig.~\ref{fig:Susc_anom}) clearly indicates anomalous magnetic entropy changes associated with the onset of long range AFM order. We conclude that,
in zero magnetic field, the magnetic entropy changes at $T_{\textrm{N}}$ are small compared to the phononic ones. In order to highlight this, we have
subtracted an arbitrary polynomial background function \textsl{f}(\textsl{T}) = a\textsl{T} + b\textsl{T}$^{3}$ from the experimental data (cf.,
e.g., \cite{klingeler2002}). The resulting data $\Delta c_{\textrm{p}}$ are shown in the bottom panel of Fig. \ref{fig:Heat_cap}. At a temperature
very close to $T_{{\rm N}}$ derived from the susceptibility measurements the data indicate a kink in $\Delta c_{\textrm{p}}$, at $B=0$, which is
associated with a specific heat anomaly $\Delta c_{\textrm{p}}<20$ mJmol$^{-1}$K$^{-1}$. This kink is hardly affected by small magnetic fields. To be
specific, at $B=5$\,T the kink is found at \textsl{T}$_{\textrm{max}}^{c}$(5 T)$=10.2$\,K. However, at magnetic fields $B\gtrsim5$\,T the peak
increases in amplitude and shifts to higher temperatures. The growth of the peak suggests the increase of entropy changes associated with the phase
transition, i.e. the evolution of a jump-like or a $\lambda$-like specific heat anomaly. At $B=9$\,T, the peak height of the specific heat anomaly
amounts to $\simeq115$ mJmol$^{-1}$K$^{-1}$ and the peak maximum occurs at $T_{\textrm{N}}$(9 T)$=11.9$\,K. The experimental data hence prove a field
induced increase of the magnetic entropy changes associated with the onset of long-range magnetic order. The vanishing size of the specific heat
anomaly in small magnetic fields, i.e. the absence of significant entropy changes associated with the onset of AFM order, either implies only weak
spin order below $T_{\textrm{N}}$ or the presence of strong spin correlations well above the long-range ordering temperature. The fact that
application of moderate external magnetic fields instead of suppression of AFM order yields larger entropy changes associated with the phase
transition corroborates the latter scenario. Hence, the specific heat data suggest strong AFM spin fluctuations at low fields. We note that from the
present data it is difficult to conclude if additional small shoulder-like features of the $\Delta c_{\textrm{p}}(T, B = 9\textrm{T})$ curve at $\sim
10$\,K and $\sim 13.3$\,K are of intrinsic nature or arise due to the imperfect subtraction of the nonmagnetic contributions from the total specific
heat.

\begin{figure}
\begin{centering}
\includegraphics[clip,width=0.95\columnwidth9]{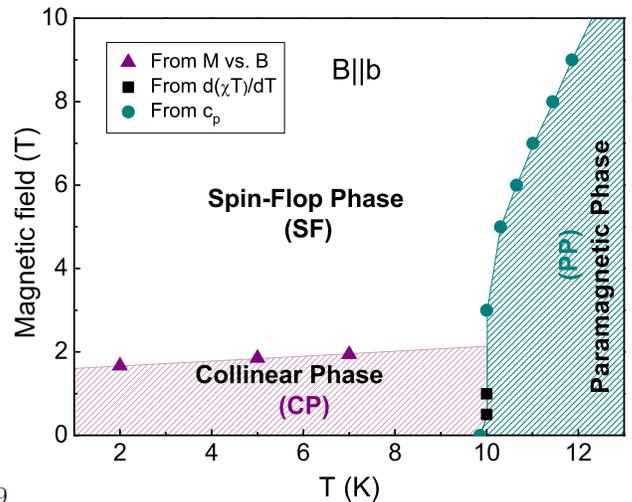}
\par\end{centering}

\caption{(Color online) Magnetic phase diagram for $\alpha$-Cu$_{2}$As$_{2}$O$_{7}$
single crystals according to magnetization and specific heat measurements
when applying field along the magnetically \textit{easy} \textit{b}-axis.}

\label{fig:Phase_dia}
\end{figure}

A tentative phase diagram which has been constructed from the magnetization and specific heat data is presented in Fig.~\ref{fig:Phase_dia}. In
particular, it shows that the spin-flop field which signals the transition from the AFM collinear phase (CP) to the spin-flop phase (SP) is nearly
independent on temperature. For a discontinuous phase transition, the slope of the phase boundary is associated with the ratio of the magnetization
changes $\Delta M$ and the entropy changes $\Delta S$ at the phase transition according to $dT/dB=-\Delta M/\Delta S$ (Clausius-Clapeyron relation
\cite{Tari}). Hence, while $\Delta M>0$ our observation of $dB/dT\simeq0$ implies the absence of significant entropy changes at the spin-flop
transition, as expected for a spin-reorientation in the AFM ordered phase. Quantitatively, at $T=1.7$\,K, we extract $dT/dB\approx 16.2$\,K/T  and
$\Delta M=2.4\cdot 10^{-3}$\,$\mu_{\rm B}$/f.u. which yields $\Delta S_{\rm SF}=0.83$\,mJ/(Mol$\cdot$K). Consquently, the degree of spin order at
high fields/low temperature in the spin-flop phase is very similar to that at zero field/low temperature. Applying the above relation at $T=5$\,K and
$T=7$\,K yields decreasing entropy changes at the spin flop transition $\Delta S_{\rm SF}(5 {\rm K})=0.64$\,mJ/(Mol$\cdot$K) and $\Delta S_{\rm
SF}(7{\rm K})=0.45$\,mJ/(Mol$\cdot$K) as already qualitatively visible from the rather constant slope of the phase boundary while $\Delta M$
decreases with temperature.

 In contrast, the phase boundary between the paramagnetic
and the AFM CP/SF phases implies $dT/dB\approx0$ for $B\leq5$\,T and $dT/dB>0$ for higher fields. Our data do not allow to judge the exact nature of
the specific heat anomaly and to quantitatively estimate the entropy changes or the specific heat jump associated to the phase transition. Firstly,
$dT/dB\approx0$ thermodynamically corresponds to only tiny entropy changes at $T_{N}$($B\lesssim 5$\,T) if the Ehrenfest equation\cite{Tari} (related
to a continuous phase transition) and the moderate kink in $\partial M/\partial T$ at the transition are considered. This indeed agrees to our
specific data. However, the pronounced \textit{positive} slope at the PP-SF transition appears in qualitative contrast to what is expected for an
ordinary AFM phase transition.

\subsection{Electron Spin Resonance (ESR)}

ESR at a frequency of $\nu\sim10$ GHz has been measured with a commercial
X-band spectrometer (Bruker EMX). High-field high-frequency ESR (HF-ESR)
measurements have been carried out with the home-made spectrometer
on the basis of the Millimeterwave Network Analyzer from AB Millimetr{{\'e}}
and a 15 T superconducting magnet from Oxford Instruments.\cite{Golze}
A well defined ESR signal of a Lorentzian shape associated with the
resonance of Cu$^{2+}$ ($3d^{9},S=1/2$) ions was observed in a broad
range of excitation frequencies and temperatures.

The dependence of the position of the ESR line $B_{{\rm res}}$ on the direction of magnetic field within the crystallographic \textit{bc} plane has
been studied at $\sim$10 GHz at room temperature (not shown). The $g$-factor calculated from this dependence as $g=h\nu/\mu_{{\rm B}}B_{{\rm res}}$
reveals an anisotropy typical for Cu$^{2+}$ ions in a square planar coordination:\cite{Abragam1970,Klingeler2006} $g_{\Vert}\simeq2.07$
(\textsl{$B\Vert b$} \textendash{} along the chains) and \textsl{$g_{\perp}\simeq2.18$} (\textsl{$B\Vert c$}). Here $h$ is the Planck constant and
$\mu_{{\rm B}}$ is the Bohr magneton.

Temperature dependent HF-ESR measurements were carried out at $\nu$ = 340 GHz. For magnetic fields applied along the magnetic \textit{easy} axis
(\textsl{$B\Vert b$}), the $g$-factor is almost temperature independent above $\sim$60\,K (Fig.~\ref{fig:ESR_temp}) close to a value of
$g_{\Vert}\simeq2.09$ which corresponds to $B_{{\rm res}}\simeq 11.6$\,T. A slight discrepancy to the 10 GHz data ($g_{\Vert}\simeq2.07$) is likely
due to a slight misorientation of the sample in the HF-ESR probe head. At lower temperature the $g$-factor drops with decreasing temperature
indicating the onset of short-range magnetic ordering when approaching \textsl{$T_{\textrm{N}}$}. The shift of the resonance field in this region
concomitant with the broadening of the line is apparently due to the growth of local quasi-static internal fields and the increase of the spin
correlation length as precursors of the long range magnetic order.

\begin{figure}
\begin{centering}
\includegraphics[clip,width=0.95\columnwidth]{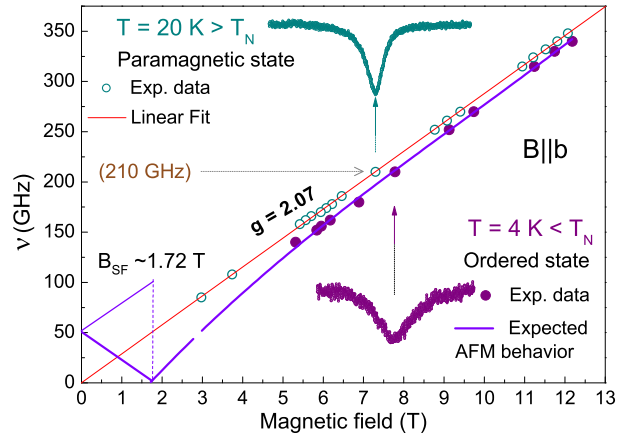}
\par\end{centering}

\caption{(Color online) Frequency versus resonance field diagram of the ESR signal in the ordered and paramagnetic states for $B\parallel b$. At 4\,K
a single resonance mode is observed above $\sim$5 T (solid circles). Solid thin line corresponds to the paramagnetic resonance branch $\nu=g\mu_{{\rm
B}}B_{{\rm res}}/h$. Solid thick lines show the expected AFM resonance branches for uniaxial anisotropy with $B\parallel$\,\textit{easy
b-}axis.\cite{Turov1965}}

\label{fig:ESR_freq}
\end{figure}

Frequency dependent measurements at $T=20$\,K$\,>\, T_{{\rm N}}$ show a linear scaling between $\nu$ and $B$ (Fig.~\ref{fig:ESR_freq}) yielding a
$g$-factor $g\simeq2.07$ from the slope $\partial\nu/\partial B$. Extrapolation of the data to zero field reveals no energy gap for the resonant
excitation as expected for an $S=1/2$ system in the paramagnetic state. An opening of a gap at $B=0$ is however expected in the magnetically ordered
state due to the magnetic anisotropy. For a simple two-sublattice collinear antiferromagnet the anisotropy gap $\Delta( B = 0)$ is related to the
spin-flop field $B_{{\rm sf}}$ as $\Delta=B_{{\rm sf}}g\mu_{{\rm B}}/h$.\cite{Turov1965} With $B_{\textrm{SF}}\simeq1.7$\,T one obtains $\Delta =
50$\,GHz. By applying the field along the \textit{easy} \textit{b}-axis two resonance modes should develop as sketched in Fig. \ref{fig:ESR_freq},
which collapse at $B_{{\rm SF}}$ \cite{Turov1965}. At $B>B_{{\rm SF}}$ one resonance branch is expected which approaches the paramagnetic $\nu(B)$
line in strong fields, as shown in Fig. \ref{fig:ESR_freq}. Indeed at high fields a well defined resonance mode has been detected at $T=4$\,K$\,<\,
T_{{\rm N}}$ which follows closely the predicted field dependence (Fig. \ref{fig:ESR_freq}). Surprisingly it fades by approaching the field of
$\sim5$\,T and no resonance signals could be found at smaller fields. Such peculiar absence of resonance modes in the low field regime is in a
remarkable correspondence with the vanishing of the specific heat anomaly below $\sim 5$\,T which again points at the pronounced spin fluctuations in
the magnetically ordered state which can be suppressed by magnetic field.

\begin{center}
\begin{figure}
\begin{centering}
\includegraphics[clip,width=0.95\columnwidth]{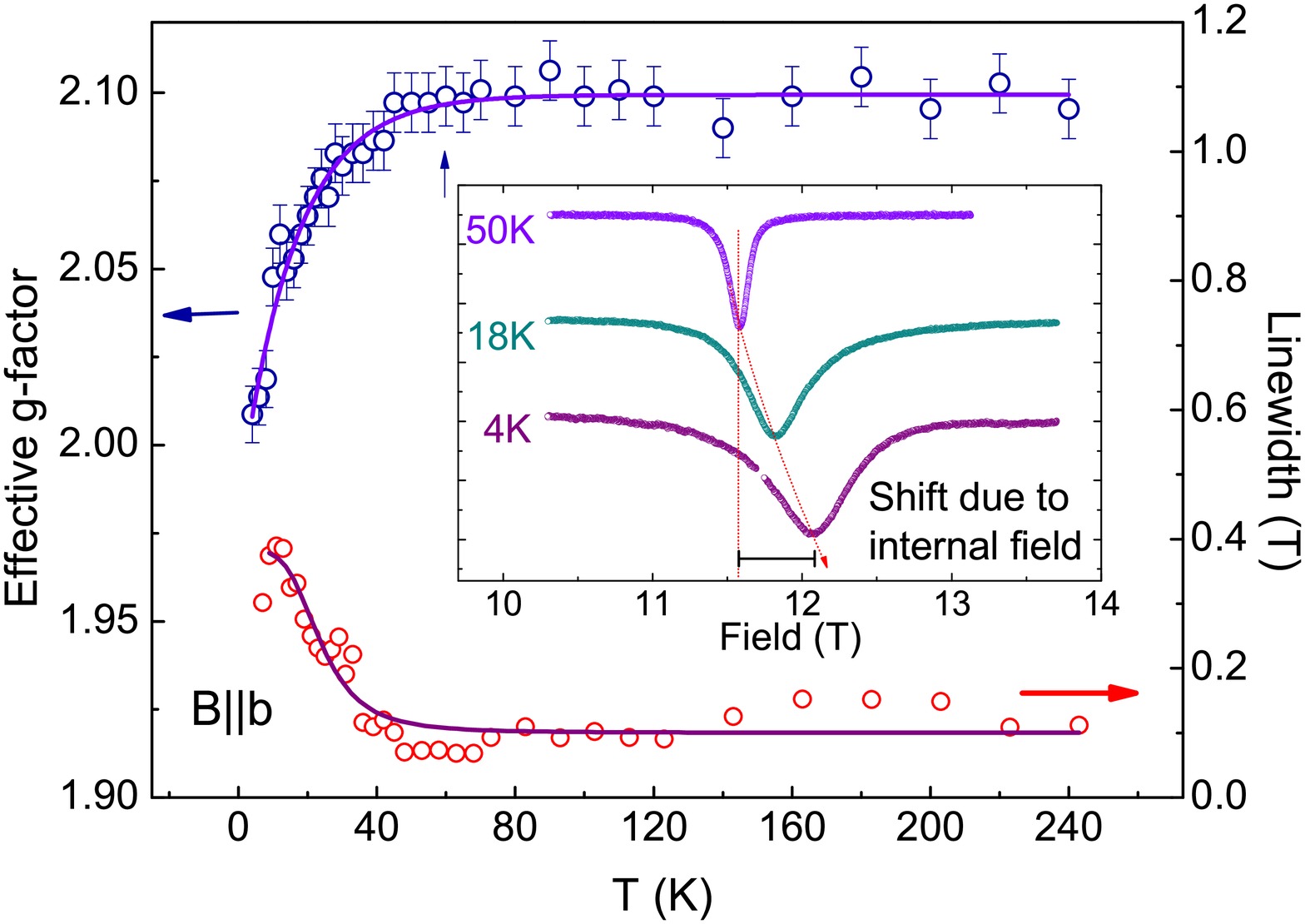}
\par\end{centering}

\caption{(Color online) Temperature dependence of the ESR parameters with $B\Vert b$
at 340 GHz. Main panel: effective $g$-factor and linewidth, upper
and lower curve respectively. Inset: The shift of the ESR resonance
line at low temperature reflects the increasing internal magnetic
field in the vicinity of the ordered state.}

\label{fig:ESR_temp}
\end{figure}

\par\end{center}

\subsection{Nuclear Magnetic Resonance (NMR)}

NMR on $^{75}$As nuclei was measured with a Tecmag pulse solid-state NMR spectrometer with a 9.2\,T superconducting magnet from Magnex Scientific.
The NMR spectra were obtained by measuring the intensity of the Hahn echo versus magnetic field. The $T_{1}$ relaxation time was measured with the
method of stimulated echo.

\begin{figure}
\begin{centering}
\includegraphics[clip,width=0.97\columnwidth]{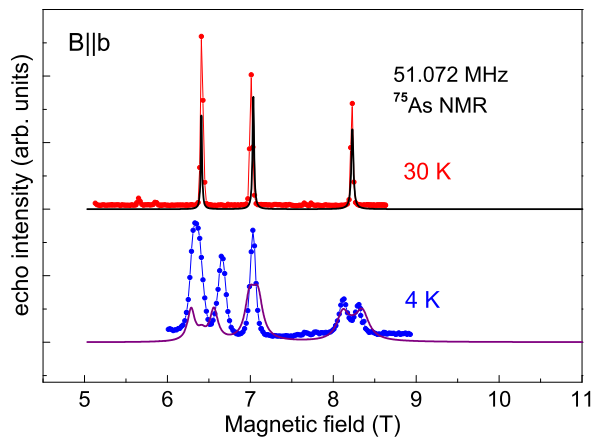}
\end{centering}
\caption{(Color online) $^{75}$As NMR spectra (dots) measured at 30 K and 4 K for the magnetic field direction parallel to the $b$-axis. Solid lines
represent simulated spectra (see the text). }

\label{NMRspectra}
\end{figure}

\begin{figure}
\begin{centering}
\includegraphics[clip,width=0.95\columnwidth]{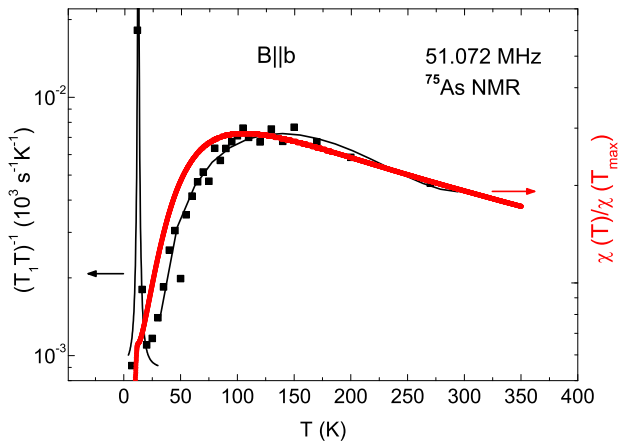}
\end{centering}
\caption{(Color online) Temperature dependence of the $^{75}$As spin lattice relaxation rate $(T_{1}^{-1}T)^{-1}$ in comparison with the static
susceptibility. Thin solid line is the guide for the eye. }

\label{NMRT1}
\end{figure}

Representative $^{75}$As NMR spectra for the \textit{easy} axis $b$-direction of the magnetic field are shown in Fig.~\ref{NMRspectra}. The
temperature dependence of the $^{75}$As spin lattice relaxation rate $T_{1}^{-1}$ divided over $T$ is shown in Fig.~\ref{NMRT1}. On this figure it is
compared with the $T$-dependence of the static susceptibility $\chi(T)$. The fact that both $(T_{1}T)^{-1}$ and $\chi(T)$ follow closely the same
temperature behavior suggests an appreciable coupling of the $^{75}$As nuclei to the Cu spin system. In this case the nuclear spin lattice relaxation
is usually governed by magnetic fluctuations in the electronic spin system and $(T_{1}T)^{-1}$ is proportional to the imaginary part of the dynamic
susceptibility $\chi''$: \begin{equation}
(T_{1}T)^{-1}\propto\sum_{\vec{q}}A_{\bot}^{2}(\vec{q},\omega)\cdot\frac{\chi''(\vec{q},\omega)}{\omega}\end{equation}
 Here $A$ is the hyperfine constant, $q$ is the wave vector and
$\omega$ is the NMR frequency. Note a sharp peak of $(T_{1}T)^{-1}$ at the ordering temperature $T_{\textrm{N}}$ manifesting a critical behavior of
Cu spin fluctuations (Fig.~\ref{NMRT1}).
At $T>T_{{\rm N}}$ the $^{75}$As NMR spectrum consists of three quadrupole-split lines as expected for nuclei with the spin $I=3/2$
(Fig.~\ref{NMRspectra}). The spectrum has been modeled on the basis of the following Hamiltonian:\cite{Abragam1961}
 \begin{equation}
\mathcal{H}=\frac{1}{6}h\nu_{{\rm Q}}^{z}\left[3I_{z}^{2}-I(I+1)+\frac{1}{2}\eta(I_{+}^{2}+I_{-}^{2})\right]+\gamma\hbar{\bf I\cdot
B}.\label{NMRham}\end{equation}
The first term here describes the quadrupole interaction and the second term the Zeeman interaction. A good agreement with the experiment has been
achieved with the value of the quadrupole frequency $\nu_{{\rm Q}}=23$ MHz and the asymmetry parameter $\eta=0.1$. In the magnetically ordered state
($T<T_{{\rm N}}$) the spectral lines experience a splitting into two components suggesting the occurrence of two nonequivalent magnetic environments
of the $^{75}$As nuclei. This can be related to the occurrence of two spin sublattices in the ordered Cu chains which exert alternating in sign
internal field at the As sites. The low temperature NMR spectra can be successfully modeled on the basis of Hamiltonian (\ref{NMRham}) with the same
values of $\nu_{{\rm Q}}$ and $\eta$ assuming two magnetically nonequivalent As sites (Fig. \ref{NMRspectra}). The modelling yields the magnitude of
the internal field at the $^{75}$As amounting to $B_{{\rm int}}=0.3$\,T which is directed approximately perpendicular to the $b$-axis. The
substantially large value of $B_{{\rm int}}$ suggests a transferred nature of the internal field due to the overlap of electronic orbitals.

The finding that $B_{\textrm{int}}\perp b$ enables a conclusion that though the NMR has been measured at fields $\sim7$ T substantially stronger that
the spin-flop field $B_{{\rm SF}}\simeq1.7$ T the Cu spins in the chains are still oriented predominantly perpendicular to the \textit{easy} axis.
This suggests the AFM exchange interaction between the Cu spins to be much stronger compared to the Zeeman energy which renders the spin canting in a
field of $\sim7$ T very small. This complies with the small magnetization values shown in Fig.~\ref{spinflop}.

\subsection{Density functional band structure calculations}

DFT calculations have been performed using the full-potential code \textsc{fplo-8.50-32}.\cite{FPLO} For the exchange and correlation potential, the
scalar-relativistic parameterization of Perdew and Wang has been used.\cite{PW92} The \textit{k}-mesh of 1152 points (320 in the irreducible wedge)
has been adopted for the nonmagnetic calculations. The transfer integrals $t_{i}$ have been evaluated based on a one-orbital tight-binding model,
parameterized using Wannier functions. Strong electronic correlations are accounted for on a model level by resorting to a Hubbard model. At
half-filling, the low-energy (magnetic) sector of this model is effectively described by a Heisenberg model.\cite{superexchange} Antiferromagnetic
exchange integrals $J_{i}^{\textrm{AFM}}$ have been determined based on transfer integrals $t_{i}$, adopting the second-order perturbation theory
expression $J_{i}^{\textrm{AFM}}=4t_{i}^{2}/U_{\textrm{eff}}$. Alternatively, the local (spin) density approximation {[}L(S)DA{]}+$U$ has been used
to evaluate the values of total exchange, comprising AFM as well as ferromagnetic (FM) contributions:
$J_{i}=J_{i}^{\textrm{AFM}}+J_{i}^{\textrm{FM}}$. The \textit{k}-meshes used for LSDA+$U$ supercell calculations number 12, 18 or 32 points,
depending on the supercell size. The results were accurately checked for convergence.

\begin{figure}[t]
 \begin{centering}
\includegraphics[angle=270,width=8.6cm]{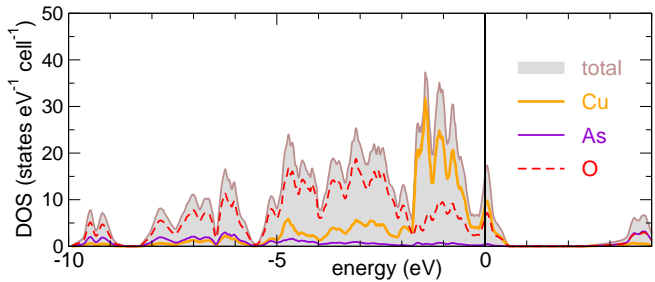}
\par\end{centering}

\caption{\label{fig:dft_LDA}(Color online) LDA total and atomic-resolved density
of states of $\alpha$-Cu$_{2}$As$_{2}$O$_{7}$. The main contribution
to the valence band is due to Cu and O states (full orange and dashed
red line, respectively). The Fermi level is at zero energy.}

\end{figure}

Typical for cuprates, the valence band of $\alpha$-Cu$_{2}$As$_{2}$O$_{7}$ is dominated by Cu and O states (Fig. \ref{fig:dft_LDA}). The nonzero
density at the Fermi level evidences a metallic ground state, in contrast to the olive color of $\alpha$-Cu$_{2}$As$_{2}$O$_{7}$ crystals. This
discrepancy, typical for the LDA, originates from insufficient treatment of strong electronic correlations, intrinsic for the $3d^{9}$ electronic
configuration of Cu$^{2+}$. Despite the inappropriate treatment of electron interaction terms, LDA correctly accounts for kinetic (exchange)
processes, and thus can be used to derive the relevant orbitals and couplings. This way, we find that the magnetic properties of
$\alpha$-Cu$_{2}$As$_{2}$O$_{7}$ are governed by a Cu $3d$
orbital of the $x^{2}-y^{2}$ symmetry.%
\footnote{The \textit{x}-axis of this local coordinate system runs along one
of the Cu--O bonds, the \textit{z} axis is perpendicular to the plaquette
plane. %
} For these states, the Wannier functions are constructed~\cite{*[{See }] [{ for description of the localization procedure.}] FPLO_WF}.

\begin{figure}
 \begin{centering}
\includegraphics[width=8.6cm]{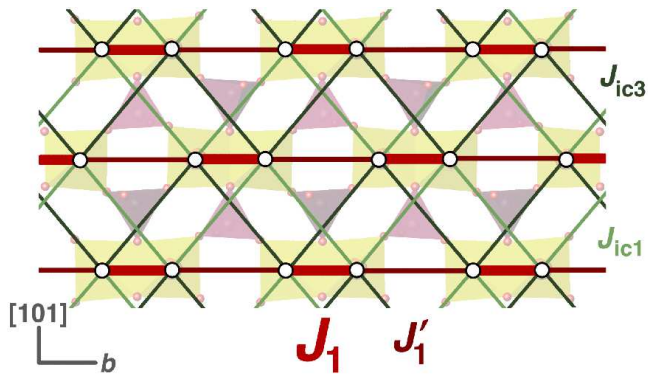}
\par\end{centering}

\caption{\label{fig:dft_microsc}(Color online) Microscopic spin model for
$\alpha$-Cu$_{2}$As$_{2}$O$_{7}$. Magnetic layers are formed by
alternating $J_{1}-J_{1}^{\prime}$ chains running along \textit{b}
that are coupled by two inequivalent couplings $J_{\textrm{ic1}}$
and $J_{\textrm{ic3}}$. The leading coupling $J_{1}$ corresponds
to the magnetic exchange within the structural Cu$_{2}$O$_{6}$ dimers. }

\end{figure}

The analysis based on Wannier functions yields four relevant (> 30 meV)
transfer integrals. The leading terms are intra- and interdimer
couplings $t_{1}$ and $t_{1}^{\prime}$, forming alternating chains
running along \textit{b}.  In addition, the chains are coupled together
by two inequivalent transfer integrals $t_{\textrm{ic1}}$ and
$t_{\textrm{ic3}}$ shaping a 2D spin lattice (Fig.
\ref{fig:dft_microsc}). Numerical values of the relevant $t_{i}$ terms
are given in Table \ref{tab:Leading-couplings-in} (third column).

Mapping onto a Hubbard model and subsequently onto a Heisenberg model
yields the values of $J_{i}^{\textrm{AFM}}$ (Table\textcolor{blue}{{}
\ref{tab:Leading-couplings-in}}). Considering $T_{\textrm{max}}=105$
K from our $\chi(T)$ measurements, the large $J_{1}^{\textrm{AFM}}=298$
K resulting from a typical value $U_{\textrm{eff}}=4.5$ eV hints
at a substantial FM contribution to this exchange. To evaluate the
values of the total exchange $J_{i}$, we perform LSDA+$U$ calculations.
This calculational scheme adds correlations on top of the L(S)DA in
a mean-field way. The implementation of the method relies on several
parameters, namely, the on-site intra-orbital term $U_{d}$, the on-site
inter-orbital term $J_{d}$, and the particular way of accounting
for the part of correlation energy already present in LDA. The latter
term is typically referred as the double-counting correction (DCC),
and two most widely used schemes are the around-mean-field (AMF),
which uses a uniform orbital occupation as a reference, and the fully
localized limit (FLL), referring to integer (0 or 1) occupation numbers.\cite{LSDA+U_Pickett}
For $U_{d}$, we used the range 6.5$\pm$1 eV within AMF and 8.5$\pm$1
eV within FLL, keeping $J_{d}=1$ eV constant.%

\begin{table}
\begin{ruledtabular}
\caption{ Leading couplings in $\alpha$-Cu$_{2}$As$_{2}$O$_{7}$: Cu--Cu distance \textit{d} (in $\textrm{\AA}$), transfer integrals $t_{i}$ (in meV) and
AFM exchange integrals $J_{i}^{\textrm{AFM}}$ (in K), calculated adopting $U_{\textrm{eff}}=4.5$ eV. Total exchange integrals $J_{i}$ (in K) are
evaluated from LSDA+$U$ calculations adopting AMF ($U_{d}=6.5$ eV) and FLL ($U_{d}=9.0$ eV) DCC. The paths are depicted in Fig.
\ref{fig:dft_microsc}.} \label{tab:Leading-couplings-in}
\begin{tabular}{ccccccc}
path & $d$ & $t_i$ & $J^{\text{AFM}}_i$ & $J_i$, AMF & $J_i$, FLL \\
\hline
$J_1$ & 3.086 & 170 & 298 & 167 & 188 \\
$J_1'$ & 5.239 & 104 & 112 & 125 & 102 \\
$J_{\text{ic}1}$ & 6.437 & 76 & 60 & 44 & 42 \\
$J_{\text{ic}3}$ & 6.465 & 77 & 61 & 46 & 42\\
\end{tabular}
\end{ruledtabular}
\end{table}

 Representative values of the resulting $J_{i}$ are given
in Table\textcolor{blue}{{} \ref{tab:Leading-couplings-in}}. %
\footnote{ The ratios of the leading couplings in the final model are mostly independent of the particular choice of $U_{d}$ and DCC, although
differences in the values of individual exchange couplings are of
the order of 30 \%.%
} In agreement with the Goodenough--Kanamori rules,\cite{GKA_1,GKA_2}
the Cu--O--Cu angle of $101.7^{\circ}$ leads to sizable FM contribution,
and consequently, to a substantial reduction of $J_{1}$. However,
in contrast to the isostructural compounds $\alpha$-Cu$_{2}$P$_{2}$O$_{7}$
(Ref.~\onlinecite{Janson}) and $\beta$-Cu$_{2}$V$_{2}$O$_{7}$
(Ref.~\onlinecite{Tsirlin2010}), this FM contribution does not suffice
to reduce $J_{1}$ strongly, thus it remains the leading coupling
in $\alpha$-Cu$_{2}$As$_{2}$O$_{7}$. FM contributions to other
relevant couplings are substantially smaller due to their long-range
nature.

 The interchain coupling topology in $\alpha$-Cu$_{2}$As$_{2}$O$_{7}$
is similar to $\alpha$-Cu$_{2}$P$_{2}$O$_{7}$, but contrasts to
$\beta$-Cu$_{2}$V$_{2}$O$_{7}$. As discussed in Ref.~\onlinecite{Janson}
, this essential difference originates from the nature of the atom
residing in the center of anionic tetrahedra, in particular, whether
it is a \textit{p}-element (P, As) or a \textit{d} element (V). The
former case favors the Cu--O--O--Cu connections via single tetrahedra,
i.e., $J_{\textrm{ic1}}$ and $J_{\textrm{ic3}}$.

 To summarize the results of the band structure calculations,
the microscopic magnetic model of $\alpha$-Cu$_{2}$As$_{2}$O$_{7}$
features alternating spin chains, with both nearest-neighbor couplings
AFM, and the dominant coupling $J_{1}$ within the structural Cu$_{2}$O$_{6}$
dimers. The chains are coupled by two inequivalent interchain couplings,
forming a 2D nonfrustrated model.

\subsection{ QMC model studies}

 DFT calculations typically provide numerically accurate
spin models, with surprisingly small errors for the individual exchange integrals $J_{i}$. Aiming to evaluate a quantitative microscopic magnetic
model with even higher precision, we perform QMC simulations of the observed quantities and subsequently compare them to the experiments.\cite{*[{For
instructive examples, see }] [{}] CuClLaNb2O7_DFT_simul_better_str, *NOCuNO33}We first attempt to describe the magnetic susceptibility of
$\alpha$-Cu$_{2}$As$_{2}$O$_{7}$ for $B\Vert b$ using the parameterized solution\cite{*[{Parameters are given in Table II of }] [{}] HC_AHC_Johnston}
for the alternating Heisenberg chain, valid in the temperature range $0.1J_{1}k_{\textrm{B}}^{-1}\leq T\leq4J_{1}k_{\textrm{B}}^{-1}$. The fit
exhibits good agreement with the experimental curve down to $\sim40$ K, yielding $J_{1}=174$ K and $J_{1}^{\prime}=112$ K that conform to our
LSDA+$U$ results. However, the fitted value of $g$-factor amounts to only 1.94, strongly underestimating the intrinsic value of 2.07 derived from
ESR.

 Next, we take the interchain couplings into account. This
way, we calculated the reduced magnetic susceptibility $\chi^{*}$
using QMC simulations of the spin Hamiltonian, by considering finite
lattices of $N=2048$ sites (32 coupled chains of 64 spins each) with
periodic boundary conditions. We used the stochastic series expansion
algorithm implemented in the code \textsc{loop}\cite{loop} from
the software package \textsc{alps}.\cite{ALPS} Low statistical errors
(below 0.1 \%) are ensured by using 50\,000 sweeps for thermalization
and 500\,000 steps after thermalization. The resulting simulated
$\chi^{*}(T^{*})$ dependencies were fitted to the experimental $\chi(T)$
curve using the expression:\begin{equation}
\chi(T)=\frac{N_{\textrm{A}}g^{2}\mu_{\textrm{B}}^{2}}{k_{\textrm{B}}J_{1}}\cdot\chi^{*}\left(\frac{T}{k_{\textrm{B}}J_{1}}\right)+\chi_{0},\label{eq:1}\end{equation}
where $\chi_{0}$ is the temperature-independent contribution, and
$N_{\textrm{A}}$ and $\mu_{\textrm{B}}$ are the Avogadro constant
and the Bohr magneton, respectively.

 As a starting point, we adopt the ratios $J_{1}:J_{1}^{\prime}:J_{\textrm{ic1}}=1:0.75:0.25$
from the LSDA+$U$ AMF calculations (Table \ref{tab:Leading-couplings-in}, last-but-one column). Although the resulting fit reasonably agrees with the
experimental curve, small deviations evidence notable differences between our microscopic model and the experimental data. To adjust the model
parameters, we varied $J_{1}^{\prime}/J_{1}$ and $J_{\textrm{ic1}}/J_{1}$ ratios in a broad range, while the two relevant interchain couplings
$J_{\textrm{ic1}}$ and $J_{\textrm{ic3}}$ were kept equal. This way, the best fit is attained for
$J_{1}:J_{1}^{\prime}:J_{\textrm{ic1}}=1:0.55:0.20$, in almost perfect agreement with the ratios $1:0.54:0.22$ from the FLL DCC of LSDA+$U$ (Table
\ref{tab:Leading-couplings-in}, last column), with the absolute value of $J_{1}=164$ K only 13 \% smaller than its LSDA+$U$ FLL estimate. Even more
remarkable, the QMC fit yields $g$=2.06, in excellent agreement with the experimental ESR estimate of 2.07 for $B\Vert b$. Therefore, the QMC
simulations reveal crucial importance of interchain couplings for a proper description of the thermodynamic properties of
$\alpha$-Cu$_{2}$As$_{2}$O$_{7}$.
\begin{figure}[H]
 \centering{}\includegraphics[angle=270,width=8.6cm]{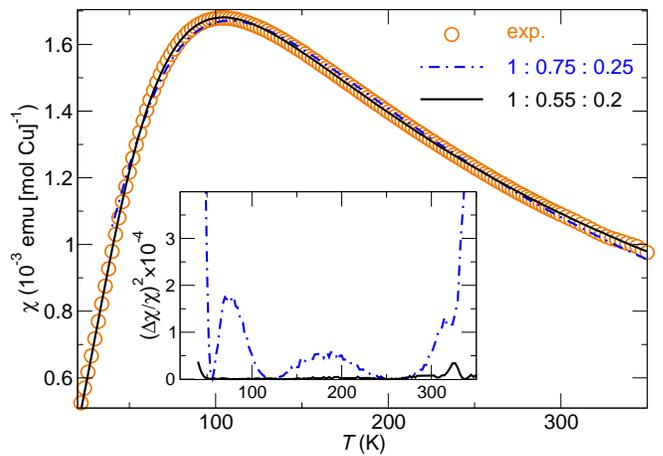}\caption{(Color online) QMC fits to the experimental (exp.) magnetic susceptibility.
The $J_{1}:J_{1}^{\prime}:J_{\textrm{ic1}}$ ratios are given in the
legend. The difference plots (inset) indicate the relative goodness
of fit $\left(\Delta\chi/\chi\right)^{2}$ for the two parameter sets
of $J_{1}:J_{1}^{\prime}:J_{\textrm{ic1}}$.}
\end{figure}

\begin{figure}[H]
\centering{}\includegraphics[angle=-90,width=\columnwidth]{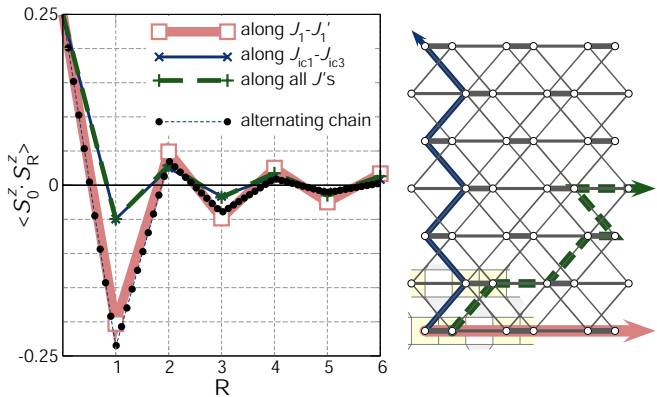} \caption{\label{F-S0SR}(Color online) Left panel: diagonal spin
correlations $\langle{}S^z_0S^z_{\text{R}}\rangle$(R) in the magnetic ground state of Cu$_2$As$_2$O$_7$. The correlations along
$J_{1}$--$J_{1}^{\prime}$ bonds (thick solid line, empty squares) closely resemble those of an alternating Heisenberg chain with
$J_{1}^{\prime}$:$J_{1}$\,=\,0.55 (dotted line). In contrast, the correlations between the $J_{1}$--$J_{1}^{\prime}$ chains are substantially smaller
(dashed and thin solid lines). Right panel: 0--$R$ pathes in the spin lattice of Cu$_2$As$_2$O$_7$ used for the
$\langle{}S^z_0S^z_{\text{R}}\rangle$(R) plot in the left panel.} \end{figure}

To investigate the role of interchain couplings for the magnetically
ordered ground state, we calculate diagonal spin correlations
$\langle{}S^z_0S^z_{\text{R}}\rangle$ for 0\,$\leq$\,R\,$\leq$6 along
different paths on the spin lattice (Fig.~\ref{F-S0SR}, right panel).
First, we address the correlations between the alternating
$J_{1}$--$J_{1}'$ chains (Fig.~\ref{F-S0SR}, dashed and thin solid
lines). As expected, $J_{\textrm{ic1}}$ and $J_{\textrm{ic3}}$ give rise
to substantial interchain correlations, and thus favor a quasi-2D
magnetic model. However, the correlations along the
$J_{1}$--$J_{1}^{\prime}$ alternating chains (Fig.~\ref{F-S0SR}, thick
line) nearly coincide with the spin correlations of an alternating
Heisenberg chain featuring the same ratio (0.55) of the nearest-neighbor
couplings (Fig.~\ref{F-S0SR}, dotted line). Therefore, the interchain
couplings, despite their crucial role for long-range magnetic ordering,
have only a minor impact on the key element of the magnetic model ---
$J_{1}$--$J_{1}^{\prime}$ alternating spin chains.
As a result, even the purely 1D $J_1$--$J_1'$ model reasonably fits the
experimental magnetic susceptibility, although the underestimated
$g$ value and the magnetic transition at $T_{\text{N}}$ evidence notable
deviations from such a simplified model.

\subsection{ Structure-property relationships}

 Since continuous symmetries can not be spontaneously broken
at finite temperature neither in one nor in two dimensions,\cite{Mermin_Wagner}
the observed long-range AFM order below $T_{\textrm{N}}=10$ K can
be accounted for only by extending the model with an effective coupling
between the magnetic layers. Considering the small size of this coupling,
its accurate estimate based on DFT results is at best challenging.
In addition, the propagation vector of the magnetically-ordered structure
is not known, impeding a conclusive analysis of the interchain coupling
regime on a microscopic level. Therefore, the magnetic ordering temperature
as well as the propagation vector are beyond the scope of the present
study.

 Magnetic layers formed by coupled structural dimers are
a common element in spin-$\frac12$ systems based on Cu$^{2+}$ {[}$\alpha$-Cu$_{2}$P$_{2}$O$_{7}$
(Ref.~\onlinecite{Janson}) and $\beta$-Cu$_{2}$V$_{2}$O$_{7}$
(Ref.~\onlinecite{Tsirlin2010}){]} as well as V$^{4+}$ {[}(VO)$_{2}$P$_{2}$O$_{7}$
(Ref.~\onlinecite{[{For example: }][{}]AHC_VO2P2O7_chiT,*AHC_VO2P2O7_DFT}),
CsV$_{2}$O$_{5}$ (Ref.~\onlinecite{CsV2O5_DFT_1_chiT_fit,*CsV2O5_DFT_2,*AV2O5_chiT_MH}),
and VOSeO$_{3}$ (Ref.~\onlinecite{VOSeO3_chiT,*VOSeO3_DFT}){]}.
A key feature of their microscopic magnetic models is the minor role
of the coupling $J_{1}$ within structural dimers, compared to the
stronger coupling running via the double bridge of nonmagnetic anions
between the structural dimers ($J_{1}^{\prime}$ in our present notation).\cite{[{Note, particularly, the experimental evidence in }][{}]garrett1997}
The reduction in $J_{1}$ is generally understood as the suppression
of Cu--O--Cu or V--O--V superexchange due to a bond angle slightly
exceeding 90$^{\circ}$. In contrast, $\alpha$-Cu$_{2}$As$_{2}$O$_{7}$
stands out against this trend, since the dominating coupling $J_{1}$
is restored unexpectedly. The enhanced value of $J_{1}\simeq164$
K is only partially related to the change in the Cu--O--Cu bond angle
that amounts to 101.7$^{\circ}$, compared to 100.4$^{\circ}$ in
$\alpha$-Cu$_{2}$P$_{2}$O$_{7}$ ($J_{1}\simeq30$ K)\cite{Janson}
and 98.7$^{\circ}$ in $\beta$-Cu$_{2}$V$_{2}$O$_{7}$ ($J_{1}\leq12$
K).\cite{Tsirlin2010} An efficient mechanism affecting the coupling
could be the polarization of particular oxygen $2p$ orbitals by As
states, as proposed for Ge$^{4+}$ in CuGeO$_{3}$.\cite{FHC_CuGeO3_superexchange}
A microscopic investigation of other Cu$^{2+}$ arsenates will help
to clarify this issue.

 The coupling regime of $\alpha$-Cu$_{2}$As$_{2}$O$_{7}$
underscores the wide diversity of spin lattices arising from the quasi-1D
arrangement of structural dimers. Apart from the aforementioned Cu$^{2+}$
and V$^{4+}$ compounds, a similar structural feature is found in
Cu$_{2}$(PO$_{3}$)$_{2}$CH$_{2}$ (Ref. ~\onlinecite{Cu2PO32CH2_DFT_NMR_chiT_CpT_MH_simul})
and VOHPO$_{4}$\,$\cdot$\,0.5H$_{2}$O (Ref.~\onlinecite{tennant1997,*petit2004,*cao2005}).
Altogether, these compounds exhibit remarkably different ground states,
dimensionality, and even frustration regimes: compare, for example,
the spin gap in the quasi-1D (VO)$_{2}$P$_{2}$O$_{7}$ to the long-range
magnetic order in the quasi-2D $\beta$-Cu$_{2}$V$_{2}$O$_{7}$,
and further to the quasi-2D frustrated Cu$_{2}$(PO$_{3}$)$_{2}$CH$_{2}$
that again features a singlet ground state. $\alpha$-Cu$_{2}$As$_{2}$O$_{7}$
takes its own distinct position in this series as a quasi-1D nonfrustrated
system exhibiting a magnetic ordering transition at the notably low
$T_{\textrm{N}}/J_{1}\simeq0.06$.

\section{Conclusions}

The magnetic properties of $\alpha$-Cu$_{2}$As$_{2}$O$_{7}$ single crystals have been investigated using several experimental methods (static
susceptibility, magnetization, heat capacity, ESR and NMR measurements), as well as density-functional band-structure calculations and QMC
simulations. The results show that $\alpha$-Cu$_{2}$As$_{2}$O$_{7}$ is a spin-1/2 low-dimensional system with the crystallographic \textit{b}-axis as
the magnetically \textit{easy} axis. The intrachain interactions are reflected in 1D short range AFM correlations below $\sim$100 K and short range
magnetic ordering below $\sim$60 K. A significant interchain coupling contribution drives the system into an AFM long range order at
\textsl{$T_{\textrm{N}}\simeq10$} K. We find that a second phase transition of spin-flop type takes place at \textsl{$T<T_{\textrm{N}}$} for
\textit{$B_{\textrm{SF}}\simeq1.7$} T (2 K), most likely related to the gain in magnetic anisotropy energy. The collinear AF magnetic ordering below
\textit{B}$_{SF}$ is confirmed by a well defined $\lambda$-like anomaly in the magnetic specific heat quantity. On contrary, neither specific heat
shows the expected $\lambda$-like specific heat anomaly below $\sim$5,T, nor ESR signals are detectable in this low field range suggesting the
occurrence of strong spin fluctuations at small fields. The occurrence of the AFM resonance mode and the splitting of the $^{75}$As NMR signals
observed in the high field regime suggests the robust AFM spin order in the spin flop phase above $\sim$5\,T. The DFT calculations yield a reliable
determination of all relevant magnetic couplings. A combined analysis of experimental and theoretical data enables a conclusion that, in contrast to
other isostructural compounds, the pyroarsenate $\alpha$-Cu$_{2}$As$_{2}$O$_{7}$ can be effectively described by a\textit{ coupled} alternating
Heisenberg chain model with the leading intrachain coupling placed within the structural dimers Cu$_{2}$O$_{6}$. The topology and the strength of
exchange couplings determined from DFT calculations render a spin
frustration scenario unlikely and rather put forward low-dimensional nature of the spin
system. We conclude that indeed the electronic state of the central ion in the non-magnetic AsO$_{4}$ side groups plays a crucial role for
determining the relevant interchain pathways which makes $\alpha$-Cu$_{2}$As$_{2}$O$_{7}$ distinct in its magnetic properties from other
representatives of this class of compounds.

\begin{acknowledgements}

The work was supported in part by the DFG through Grants No. BU 887/13-1, KL 1824/2-2, and FOR912, and by the RFBR through Grant No.
08-02-91952-NNIO. Y. C. A. acknowledges support of the EU Programme Al{{\ss}}an. A. A. T. was funded by Alexander von Humboldt foundation.

\end{acknowledgements}

\bibliography{Cu2As2O7_final}

\end{document}